\documentclass[]{elsarticle}

\usepackage[utf8]{inputenc}
\usepackage{graphicx}
\usepackage{subfigure}

\usepackage{amssymb}
\usepackage{amsthm}
\usepackage{url}
\usepackage{amsmath}
\usepackage{subfigure}
\usepackage{textcomp}
\usepackage{xcolor}
\usepackage[margin=3cm]{geometry}

\graphicspath{{figs/}}

\title{On resistive spiking of fungi}

\author[1]{Andrew Adamatzky}
\author[2,1]{Alessandro Chiolerio}
\author[3,1]{Georgios Sirakoulis}

\address[1]{Unconventional Computing Laboratory, UWE, Bristol, UK}
\address[2]{Center for Sustainable Future Technologies,
Istituto Italiano di Tecnologia, Torino, Italy}
\address[3]{Department of Electrical and Computer Engineering
Democritus University of Thrace, Xanthi, Greece}

\begin{document}

\begin{frontmatter}

\begin{abstract}
We study long-term electrical resistance dynamics in mycelium and fruit bodies of oyster fungi \emph{P. ostreatus}. A nearly homogeneous sheet of mycelium on  the surface of a growth substrate exhibits trains of resistance spikes. The average width of spikes is c.~23~min and the average amplitude is c.~1~kOhm. The distance between neighbouring spikes in a train of spikes is c.~30~min. Typically there are 4-6 spikes in a train of spikes. Two types of resistance spikes trains are found in fruit bodies: low frequency and high amplitude (28~min spike width, 1.6~k$\Omega$ amplitude, 57~min distance between spikes) and high frequency and low amplitude (10~min width, 0.6~k$\Omega$ amplitude, 44~min distance between spikes). The findings could be applied in monitoring of physiological states of fungi and future development of living electronic devices and sensors.
\end{abstract}


\end{frontmatter}

\section{Introduction}

Electrical resistance of living substrates is used to identify their morphological and physiological state~\cite{crile1922electrical,schwan1956specific,mcadams1995tissue,heroux1994monitoring,dean2008electrical}. Examples include 
determination of states of organs~\cite{gersing1998impedance},
detection of decaying wood in living trees~\cite{skutt1972detection,al2006electrical},
estimation of roots vigour~\cite{taper1961estimation},
study of freeze-thaw injuries of plants~\cite{zhang1992electrical}, as well as 
classification of breast tissue~\cite{da2000classification}.
The aim of this paper is two-fold. 

First aim is to study the dynamics of the fungal resistance during long-term (up to two days of intermittent measurements). Whilst resistive properties of plants and mammals tissue have been studied extensively, results on electrical resistance of fungi are absent. This gap should be properly filled because the fungi is the largest, widely distributed and the oldest group of living organisms~\cite{carlile2001fungi}. Fungi  ``possess almost all the senses used by humans''~\cite{bahn2007sensing}: they can sense light, chemicals, gases, gravity and electric fields. Fungi show a pronounced response to changes in a substrate pH~\cite{van2002arbuscular}, demonstrate mechanosensing~\cite{kung2005possible} and sensing of toxic metals~\cite{fomina2000negative}, CO$_2$~\cite{bahn2006co2}, and chemical cues, especially stress hormones, from other species~\cite{howitz2008xenohormesis}. Thus mycelium networks can be used as large-scale distributed sensors. To prototype fungal sensing networks we should know their electrical features and resistance is definitely one of these characteristics.

Second aim is to assess whether fungi can be employed as electronic oscillators. The application domain of the fungal electronic oscillators could be the field of unconventional computing~\cite{adamatzky2016advances}, especially in the framework of organic electronics, living sensor and living computing wetware. Feasibility studies with plants~\cite{gizzie2016hybridising,gizzie2016living}, slime mould~\cite{berzina2015hybrid,romeo2015bio,berzina2018biolithography,cifarelli2014loading} and fungi~\cite{beasley2020memfractance,beasley2020capacitive} have shown that it is possible to develop electrical analog computing circuits either based or with these living creatures. Biological molecules such as suine microtubules have been shown to enable very fast oscillations, in the tents of MHz range ~\cite{MT2020}. However, to have a full functional analog computer, we probably need an oscillator. As Horowitz and Hill reported --- ``A device without an oscillator either doesn't do anything or expects to be driven by something else (which probably contains an oscillator)''~\cite{horowitz1980}. Thus, we envisage that the resistive spiking can be utilised to produce fungal electronic oscillators.

\section{Methods}
\label{methods}

\begin{figure}[!tbp]
    \centering
    \subfigure[]{\includegraphics[width=0.32\textwidth]{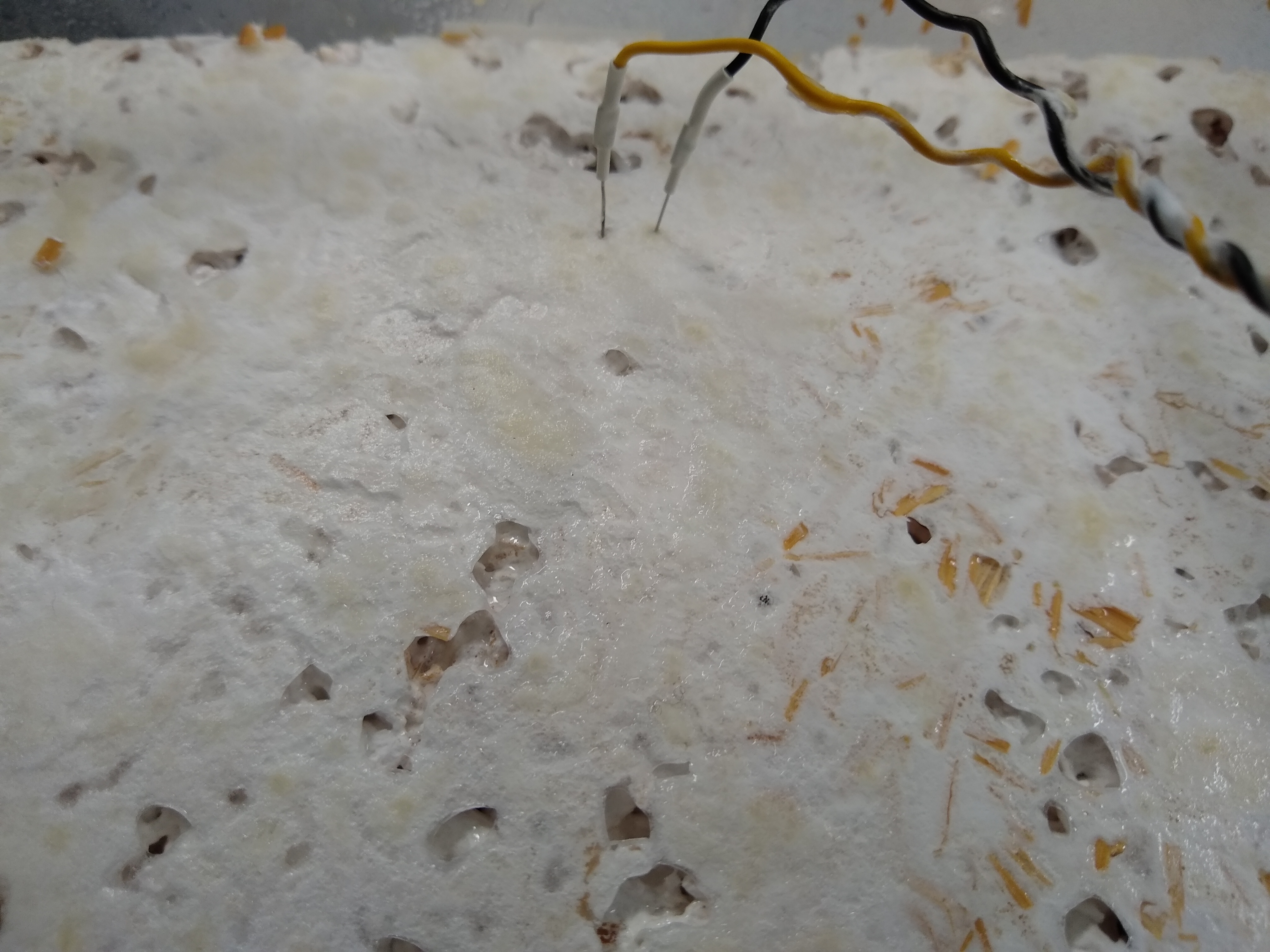}\label{fig:electrodesSubstrate}}
    \subfigure[]{\includegraphics[width=0.32\textwidth]{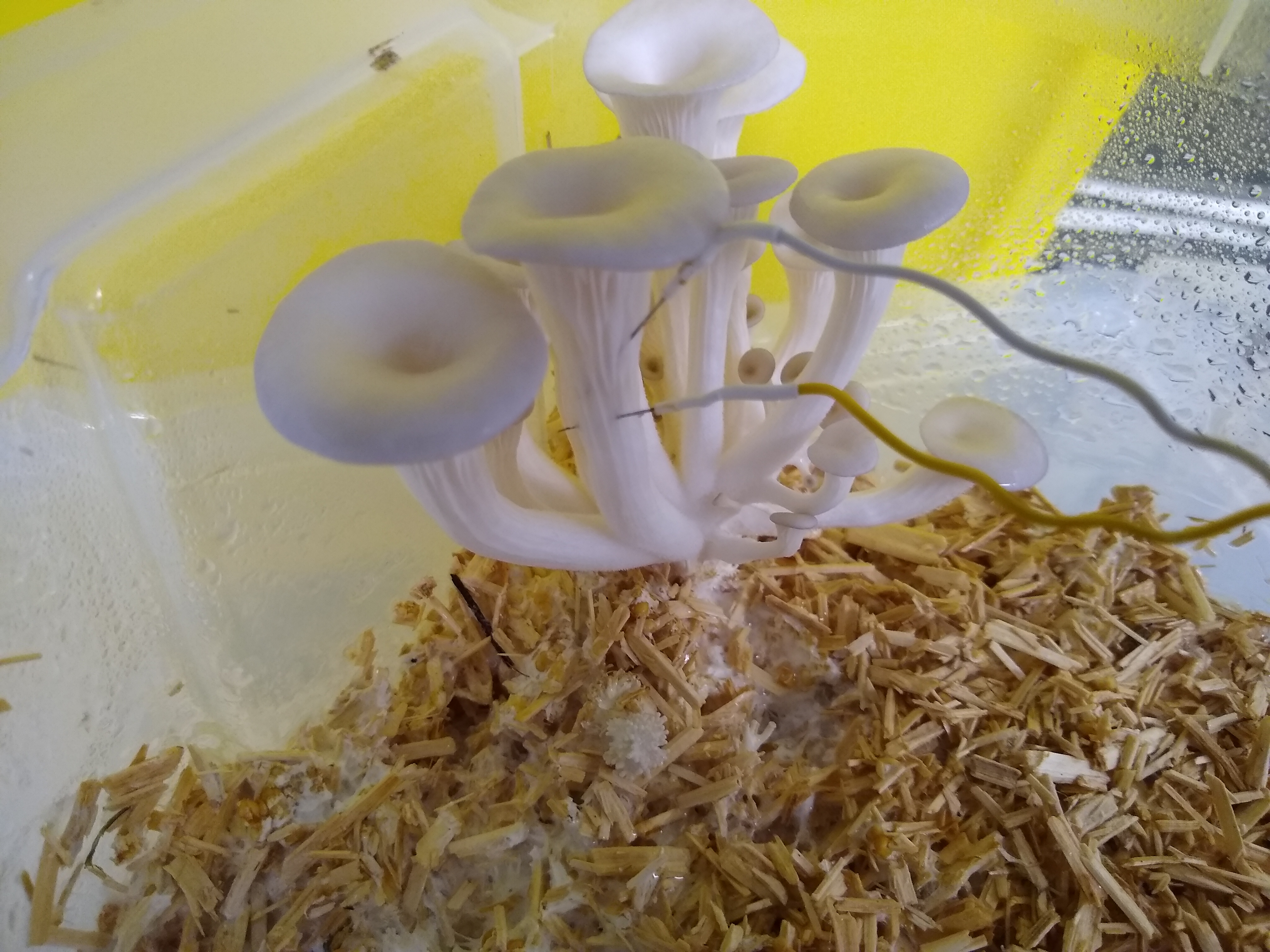}\label{fig:electrodesfruit}}
    \subfigure[]{\includegraphics[width=0.16\textwidth]{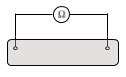}\label{fig:schemeOhm}}
    \subfigure[]{\includegraphics[width=0.16\textwidth]{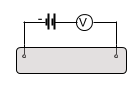}\label{fig:schemeDCVapplied}}
    \caption{Experimental setup. (a)~Electrodes are inserted in a hemp substrate nearly fully colonised by \emph{P. ostreatus}. (b)~Electrodes are inserted in stalk of a fruit body of \emph{P. ostreatus}.
    (c)~Scheme of resistance measurement. 
    (d)~Scheme of measuring fungal electrical potential under DC applied.
    }
    \label{fig:electrodes}
\end{figure}

Oyster fungi \emph{P. ostreatus} have been cultivated on hemp substrate in plastic containers in darkness and at ambient temperature 20-23~\textsuperscript{o}C. We used the substrate after it was nearly fully colonised by mycelium, which was indicated by an almost everywhere white colour and white film of nearly homogeneous mycelium, sometimes called `skin' on surface of the substrate that was formed. The electrical resistance of the skin was measured as follows. We used iridium-coated stainless steel sub-dermal needle electrodes (Spes Medica S.r.l., Italy), with twisted cables. The pairs of electrodes were inserted in fungal skin, while the distance between electrodes was kept 1~cm (Fig.~\ref{fig:electrodesSubstrate}). Twelve trials of measurements were undertaken with fungal skin. In six trials, we also undertook recordings of the fruit body's resistance, where electrodes were inserted in stalks of the bodies (Fig.~\ref{fig:electrodesfruit}). The resistance was measured (Fig.~\ref{fig:schemeOhm}) and logged using Fluke 8846A precision multimeter, the test current being 1$\pm$0.0013 $\mu$A, once per 10~sec, $5 \cdot 10^4$ samples per trial. When characterising trains of spikes, we measured spike average width $w$, average amplitude $a$ and average distance between spikes $d$. To check if there are potential oscillations of voltage applied to the fungi, we applied direct current voltage with GW Instek GPS-1850D laboratory DC power supply and measured voltage with Fluke 8846A (Fig.~\ref{fig:schemeDCVapplied}).

\section{Results}
\label{results}

\begin{figure}[!tbp]
\subfigure[]{
    \includegraphics[width=0.49\textwidth]{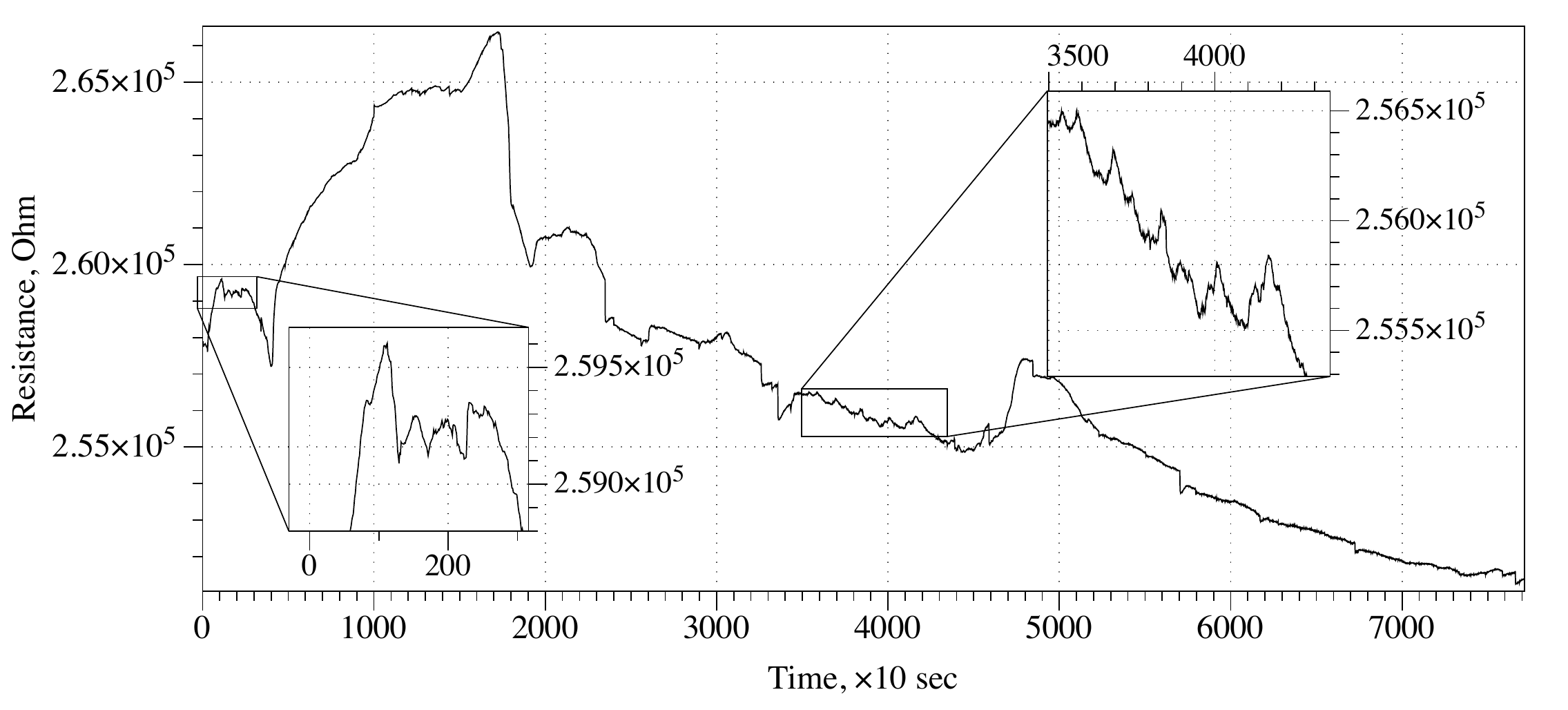}\label{fig:longtermrecording}
    }
\subfigure[]{
    \includegraphics[width=0.49\textwidth]{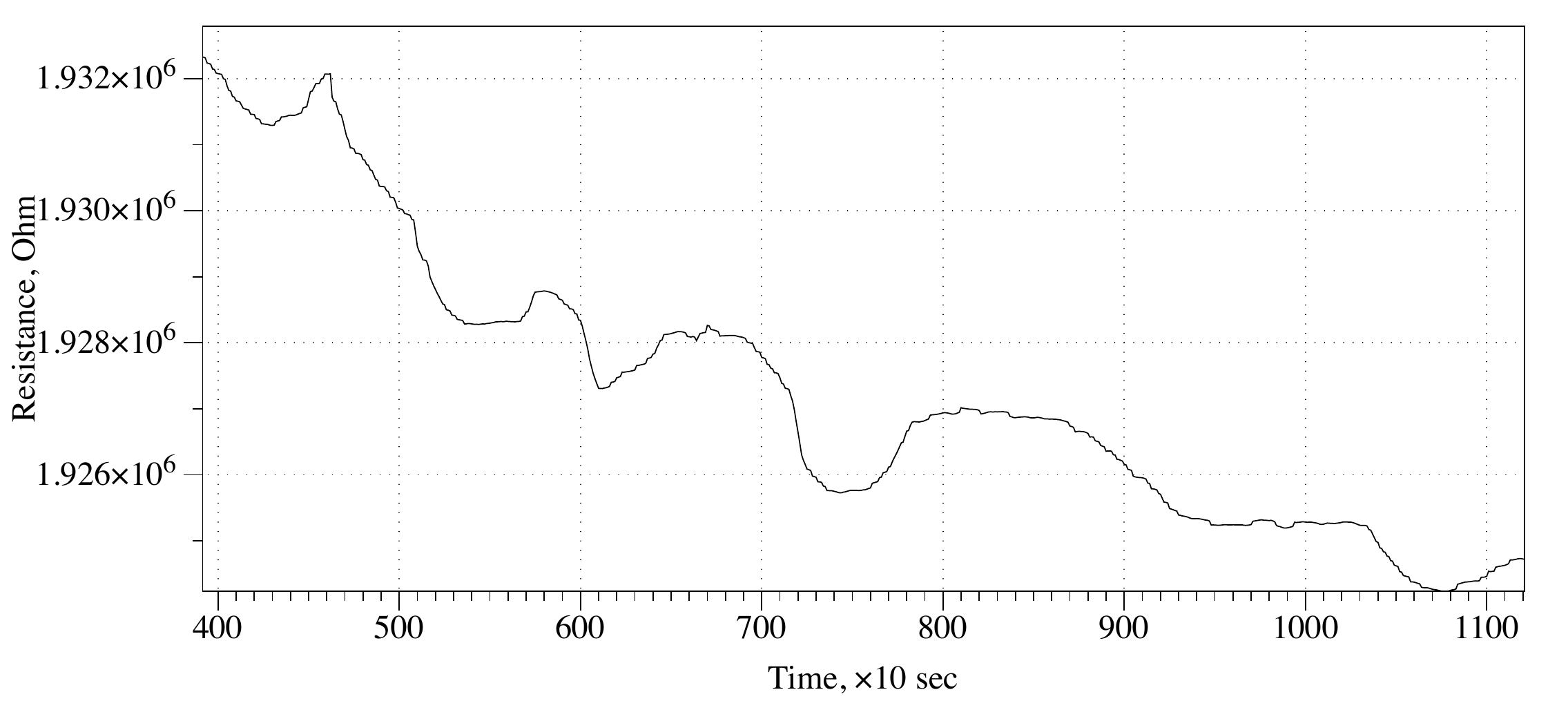}\label{fig:exampleTrain}
    }
\subfigure[]{
    \includegraphics[width=0.49\textwidth]{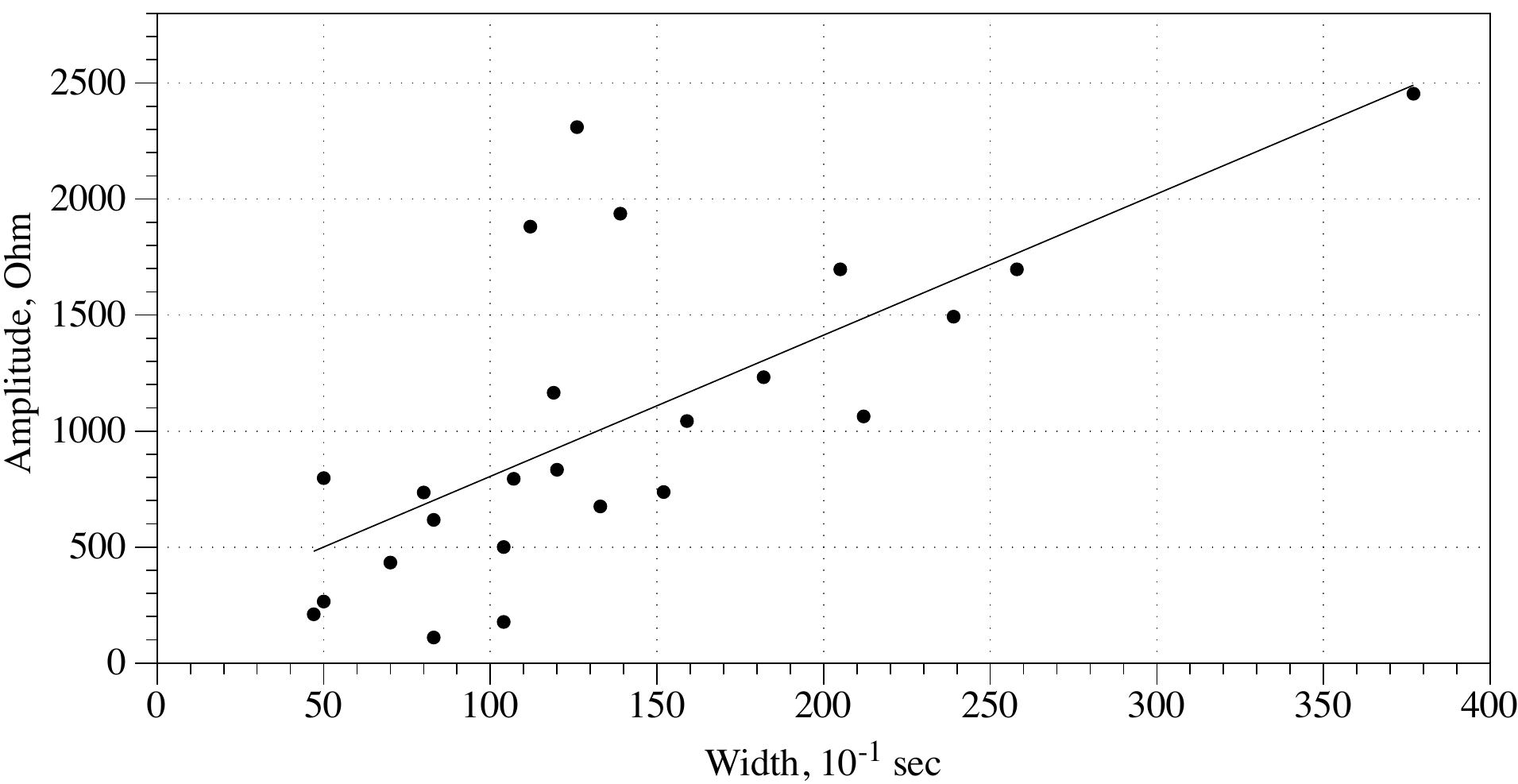}\label{fig:widthVsamplitude}}
\subfigure[]{
    \includegraphics[width=0.49\textwidth]{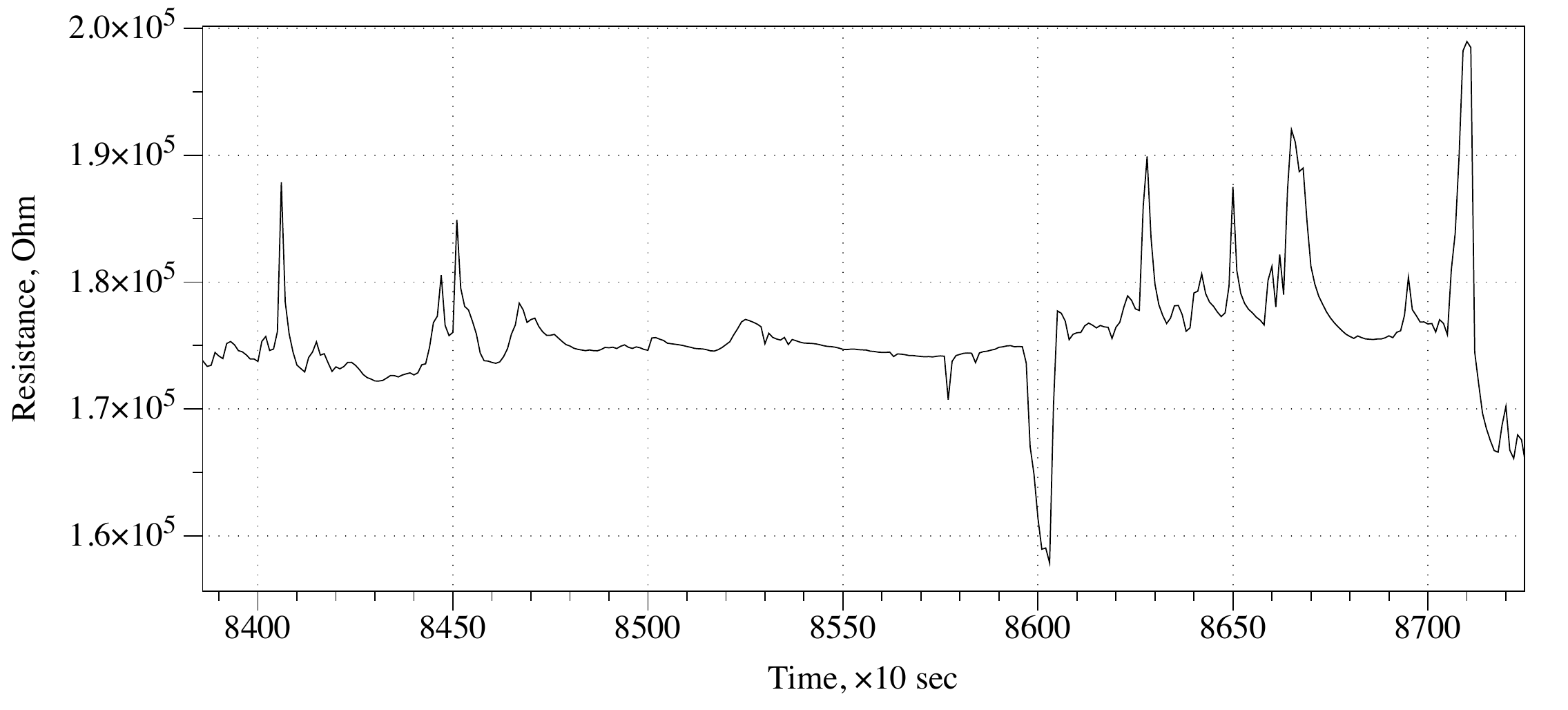}\label{fig:HighFrequency}
    }
    \subfigure[]{
    \includegraphics[width=0.49\textwidth]{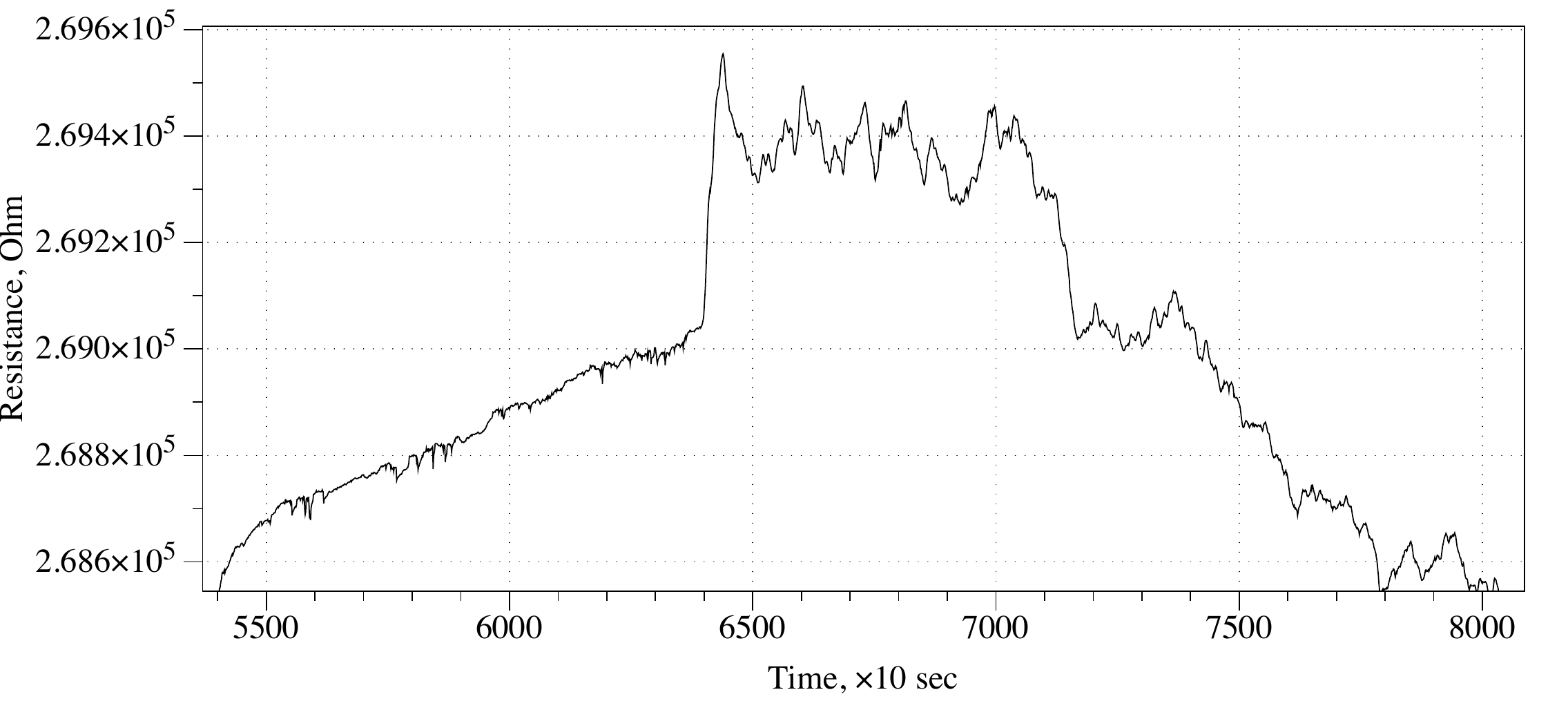}\label{fig:spikeOnHill}
    }
    \subfigure[]{
    \includegraphics[width=0.49\textwidth]{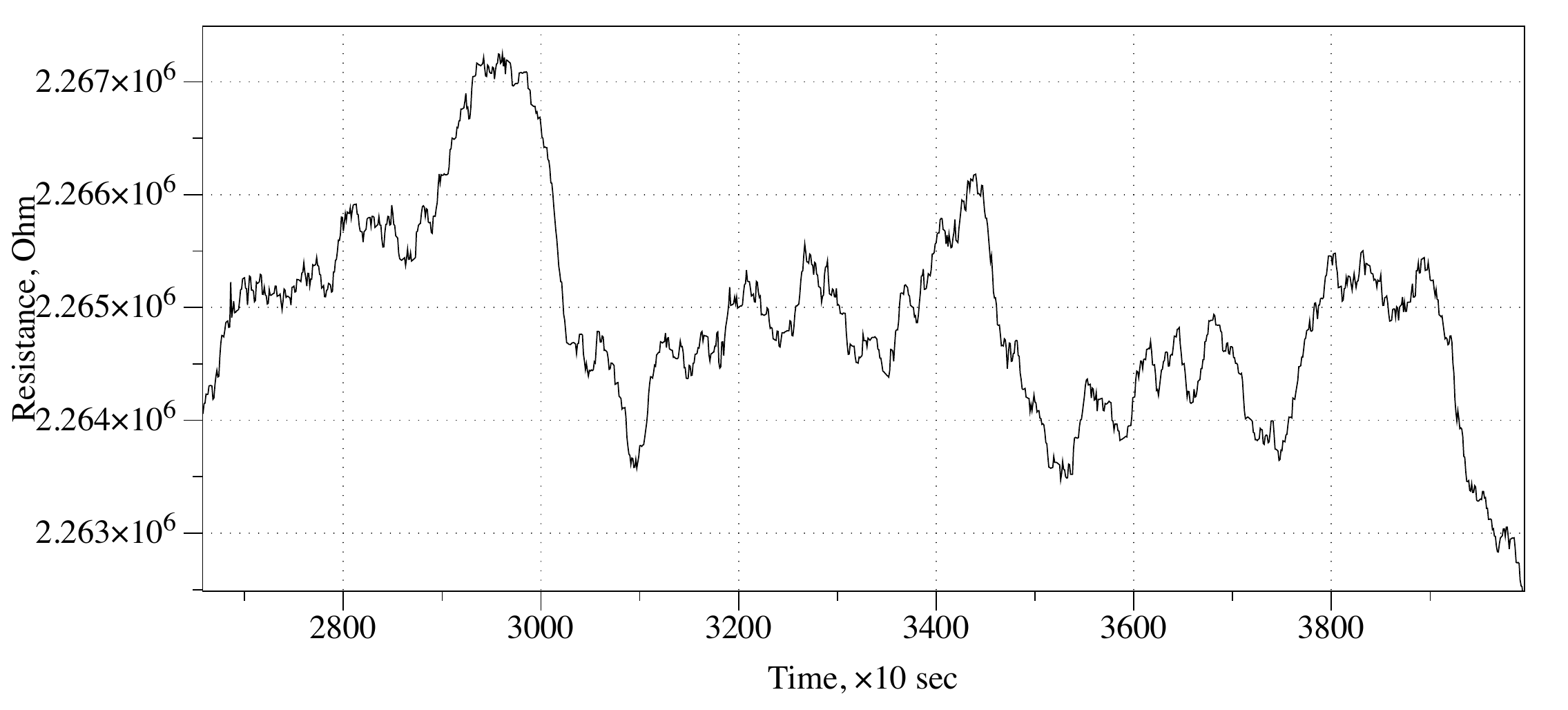}\label{fig:spikesFuitBodies}
   }
     \subfigure[]{
    \includegraphics[width=0.49\textwidth]{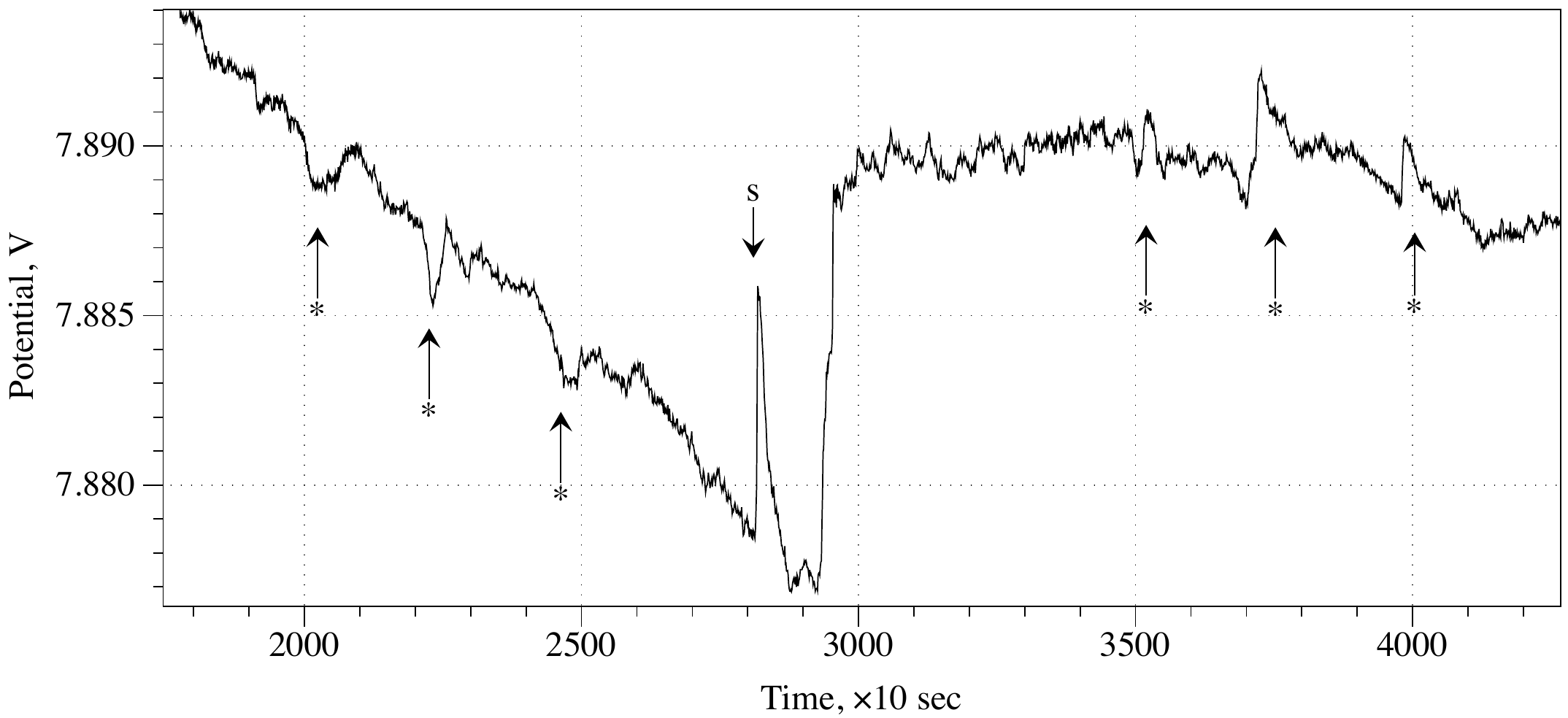}\label{fig:voltageSpikes}}
    \caption{(a)~Slow variations of resistance with trains of spikes, zoomed in the inserts, are usually observed in long term recordings.
    (b)~Example of a train of 5 spikes. 
    (c)~Spike width $w$ versus amplitude $a$ distribution. Line is a linear fit $a=6w+195$, $R^2=0.49$.
    (d)~Examples of high amplitude and high frequency spikes.
    (e)~Example of a spike train on top of a very slow variation of resistance.
    (f)~Example of resistance recorded on fruit bodies.
    (g)~Oscillation of electrical potential under 10~V DC applied, where spikes analysed are marked by `*'.}
    \label{fig:examplsOfSpikes}
\end{figure}

The resistance of fungal mycelium exhibits very slow, $1.5\cdot 10^4$--$3\cdot 10^4$~sec, disordered changes of the resistance values with trains of spikes, of increased resistance, emerging. An example of the long-term recording is shown in Fig.~\ref{fig:longtermrecording} and a train of spikes in Fig.~\ref{fig:exampleTrain}. In over 16 trials we inferred the following parameters of spikes: $w=1380$~sec (median 1190~sec, $\sigma=77$), $a=1,036\Omega$ (median 815$\Omega$, $\sigma=674$), $d=1830$~sec (median 175~sec, $\sigma=87$). Spike width versus amplitude distribution is shown in Fig.~\ref{fig:widthVsamplitude}. 

Two trials of the resistance recording from substrate, colonised by fungi, have shown outstanding phenomena (although these have not been explicitly included in the above analysis). 

More specifically, in one trial (not included in the analyses above) we observed high frequency ($w=86$~sec, median 86~sec, $\sigma=7$ and $d=283$~sec, median 250~sec, $\sigma=166$) and high amplitude ($a=11,448\Omega$, median 10,750$\Omega$, $\sigma=3,664$). An example of these high amplitude spikes is shown in  Fig.~\ref{fig:HighFrequency}.

In another trial we observed very slow variations of resistance (c. $3\cdot 10^4$ from the start of the ascent to the end of the descent). On tops of these variations there were trains of 5-7 spikes. An example is shown in Fig.~\ref{fig:spikeOnHill}. Average widths $w$ of these spike is $1,094$~sec ($\sigma=475$), $a=158\Omega$ ($\sigma=49$) and $d=1,135$~sec ($\sigma=442$). 

In fruit bodies, we typically recorded two types of spikes: low frequency ($w=1,690$~sec, $\sigma=32$ and $d=3,450$~sec, $\sigma=161$) and high amplitude ($a=1,632\Omega$, $\sigma=116$) and high frequency ($w=580$~sec, $\sigma=16$ and $d=2,630$~sec, $\sigma=188$) and low amplitude ($a=611$~Ohm,  $\sigma=266$). Figure~\ref{fig:spikesFuitBodies} shows a typical train of four high frequency spikes followed by a train of low frequency spikes. 

To assess feasibility of the living fungal oscillator, we conducted a series of scoping experiments by applying direct voltage to the fungal substrate and measuring output voltage. An example of the electrical potential of a substrate colonised by fungi under 10~V applied is shown in Fig.~\ref{fig:voltageSpikes}. Voltage spikes are clearly observed. Spikes with amplitude above 1~mV, marked by `*', except the spike marked by `s' have been analysed. We can see two trains of three spikes each. Average width of the spikes is 1,050~sec ($\sigma=9.2$, median 1,090~sec), average amplitude 2.5~mV  ($\sigma=0.68$, median 2.2~mV), while average distance between spikes is 2,318~sec ($\sigma=25.6$, median 2370~sec).

\section{Discussion}
\label{discussion}

We demonstrated that oyster fungi \emph{P. ostreatus} undergo oscillations of resistance with trains of resistive spiking emerging. Spikes amplitude vary from $1k\Omega$ to $1.6k\Omega$ and width of spikes from 23~min up to 28~min. A distance between spikes in a train varies from 30~min to nearly 60~min. The oscillations of resistance have so low frequency that could not be explained using conventional electronics framework (e.g. charging of a mycelium during probing) and resistance sampling was with very low frequency (once per 10~sec). Thus the only feasible explanation, we see is the translocation of water and metabolites taking place in the mycelium. This translocation is periodic, and more likely guided by calcium waves. Increase in a liquid in the mycelium loci leads to reduced resistance. When the translocated mass of metabolites leaves the area, the resistance increases. Rates of the translocation, measured by injecting fluoresecent dye in hyphae, reported in \cite{schutte1956translocation} are 2--6~cm/hour for small specimen and 9--15~cm/hour for large specimen. The distance between electrodes in our experiments was 1~cm, thus the above rate can be translated to the following width of resistive spikes --- 10--30~min and 4--7~min. The first estimates matches in scale with resistive spikes measured in our experiments. The widths of resistive spikes are twice the widths of electrical potential spikes observed by us previously in fruit bodies of \emph{P. ostreatus}~\cite{adamatzky2018spiking}. All the above indicate that the resistive spiking observed is not an artefact but manifestation of physiological processes in fungal mycelium and fruit bodies. Therefore one of the application domains of the proposed methodological setup and delivered results could be in monitoring physiological states of fungi: the physiological states might reflect states of ecosystems inhabited by fungi~\cite{kranabetter2009epigeous}. 
In experiments with fungal oscillator we have found that at some stages the fungal skin exhibits oscillations of the electrical potential. A width of a voltage spike is c.~18~min, which is slightly less than an average width of resistive spikes, and an average amplitude c. 2.5~mV (at 10~V DC voltage applied). The amplitude is not as high as would be expected from our previous experience with slime mould electronic oscillators~\cite{adamatzky2014slime}. This may be due to the fact that in the experiments with slime mould, the electrodes were connected by a single protoplasmic tube, so its resistance was crucial, while in the fungal skin the current can also propagate along remnants of the wet hemp substrate. A very low frequency of fungal electronic oscillators does not preclude us from considering inclusion of the oscillators in fully living or hybrid analog circuits embedded into fungal architectures~\cite{adamatzky2019fungal} and future specialised circuits and processors made from living and functionalised with nanoparticles fungi.

\section*{Acknowledgement}

This project has received funding from the European Union's Horizon 2020 research and innovation programme FET OPEN ``Challenging current thinking'' under grant agreement No 858132.


\end{document}